\documentclass[aps,pre,twocolumn,groupedaddress,showpacs]{revtex4-1}
\newcommand{\bec}[1]{\mbox{\boldmath $ #1$}}
\usepackage{graphicx}
\usepackage{bm}
\usepackage{color}

\begin{document}

\title{Turbulent thermal diffusion in strongly stratified turbulence: theory and experiments}
\author{G. Amir}
\email{guyami@post.bgu.ac.il}
\author{N. Bar}
\email{barni@post.bgu.ac.il}
\author{A. Eidelman}
\email{eidel@bgu.ac.il}
\author{T. Elperin}
\email{elperin@bgu.ac.il}
\homepage{http://www.bgu.ac.il/me/staff/tov}
\author{N. Kleeorin}
\email{nat@bgu.ac.il}
\author{I. Rogachevskii}
\email{gary@bgu.ac.il} \homepage{http://www.bgu.ac.il/~gary}
\affiliation{The Pearlstone Center for Aeronautical Engineering
Studies, Department of Mechanical Engineering,
Ben-Gurion University of the Negev, P.O.Box 653, Beer-Sheva 84105,
Israel}
\date{\today}
\begin{abstract}
Turbulent thermal diffusion is a combined effect of the temperature stratified turbulence and inertia of small particles. It causes the appearance of a non-diffusive
turbulent flux of particles in the direction of the turbulent heat flux.
This non-diffusive turbulent flux of particles is proportional
to the product of the mean particle number density and the effective velocity of
inertial particles.
The theory of this effect has been previously developed only for small temperature gradients
and small Stokes numbers (Phys. Rev. Lett. {\bf 76}, 224, 1996).
In this study a generalized theory of turbulent thermal diffusion for arbitrary temperature gradients and Stokes numbers has been developed.
The laboratory experiments in the oscillating grid turbulence and in the multi-fan produced turbulence have been performed to validate the theory of turbulent thermal diffusion in strongly stratified turbulent flows.
It has been shown that the ratio of the effective velocity of
inertial particles to the characteristic vertical turbulent velocity for large
Reynolds numbers is less than 1. The effective velocity of
inertial particles as well as the effective coefficient of turbulent thermal diffusion
increase with Stokes numbers reaching the maximum at small Stokes numbers
and decreases for larger Stokes numbers.
The effective coefficient of turbulent thermal diffusion also decreases with the mean temperature gradient. It has been demonstrated that the developed theory is in a good agreement with the results of the laboratory experiments.
\end{abstract}

\maketitle

\section{Introduction}
\label{Sect.I}

Turbulent transport of inertial particles has been a subject of many studies
due to numerous applications in geophysics and environmental sciences,
astrophysics, and various industrial applications
(see, e.g., \cite{CSA80,ZRS90,BLA97,SP06,ZA08,CST11,AR10,WA00,S03,KPE07,WA09,TB09,BE10}).
Different mechanisms of large-scale and small-scale clustering
of inertial particles have been proposed.
The large-scale clustering occurs in scales which are much
larger than the integral scale of turbulence,
while the small-scale clustering is observed
in scales which are much smaller than the integral turbulence scale.

The large-scale clustering of inertial particles in non-stratified
inhomogeneous turbulence occurs due to turbophoresis phenomenon
(see, e.g., \cite{CTT75,R83,G97,EKR98,G08,MHR17})
which is a combined effect of particle inertia and inhomogeneity of turbulence.
Turbophoresis results in appearance of the additional non-diffusive
turbulent flux of inertial particles caused by the mean particle velocity
proportional to ${\bm V}_{\rm turboph} \propto - f({\rm St},{\rm Re}) \, \bec{\nabla}
\langle {\bm u} \rangle^2$, where ${\bm u}$ is the turbulent fluid
velocity, ${\rm St} = \tau_p/\tau_\eta$ is the Stokes number, $\tau_\eta=\tau_0/{\rm Re}^{1/2}$ is the Kolmogorov time, $\tau_p = m_p / (3 \pi \rho \, \nu d)$ is the Stokes time for the small spherical particles of the diameter $d$ and mass $m_p$,
${\rm Re}= \ell_0\, u_0/\nu$ is the fluid Reynolds numbers,
and $u_0$ is the characteristic turbulent velocity at the integral scale $\ell_0$ of turbulent motions and $\nu$ is the kinematic fluid viscosity. As a result of turbophoresis
inertial particles are accumulated in the vicinity of the minimum of the turbulent
intensity.

Another example of the large-scale clustering
of inertial particles in a temperature-stratified
turbulence is a phenomenon of turbulent thermal diffusion \citep{EKR96,EKR97}
that is a combined effect of the stratified turbulence
and inertia of small particles.
This phenomenon causes the appearance of a non-diffusive
turbulent flux of particles in the direction of the turbulent heat flux, i.e.,
opposite to the mean temperature gradient.
Turbulent thermal diffusion results in accumulation of the inertial
particles in the vicinity of the mean temperature
minimum and leads to the formation of
inhomogeneous spatial distributions
of the mean particle number density.
Turbulent thermal diffusion has been intensively investigated analytically
\citep{EKR98,EKR96,EKR97,EKR00,EKRS00,EKRS01,PM02,RE05}
using different theoretical approaches, in laboratory
experiments in oscillating grid turbulence
\citep{BEE04,EEKR04,EEKMR06}
and in the multi-fan produced turbulence \citep{EEKR06}.
This effect has also been detected in direct numerical simulations \citep{HKRB12}
and in atmospheric \citep{SSEKR09}
and astrophysical turbulence \citep{H16}.

In spite of intensive studies of this phenomenon,
however, turbulent thermal diffusion
has been investigated analytically only for small Stokes numbers
and for a weak temperature stratification.
On the other hand, in laboratory experiments and
in direct numerical simulations these conditions
are not always satisfied.
The goal of the present study is to investigate the phenomenon
of turbulent thermal diffusion for arbitrary temperature gradients
and various Stokes numbers. The developed theory
is validated against the data obtained in laboratory
experiments with different sources of the
turbulence production and also against the data
obtained in the atmospheric measurements.

The paper is organized as follows.
In Sect.~\ref{Sect.II} we discuss the physics of turbulent thermal diffusion.
In Sect.~\ref{Sect.III} we develop the theory of turbulent thermal diffusion
for arbitrary stratifications and Stokes numbers.
In Sect.~\ref{Sect.IV} we validate this theory
in the laboratory experiments in oscillating grid turbulence
and in the multi-fan produced turbulence.
In this section we also discuss the validation of the
theory of turbulent thermal diffusion in the atmospheric observations.
Conclusions are drawn in Sect.~\ref{Sect.V}.

\section{Physics of turbulent thermal diffusion}
\label{Sect.II}

The mechanism of the phenomenon of turbulent
thermal diffusion of inertial particles with
material density that is much larger than
the fluid density is as follows
\citep{EKR96,EKR97}.
The particle inertia (i.e., a centrifugal effect) results in a drift out of
particles inside the turbulent eddies to
the boundary regions between eddies. In these
regions the fluid pressure fluctuations as well as strain rate are maximum.
On the other hand, there is an outflow of inertial
particles from regions with minimum fluid
pressure fluctuations (maximum vorticity).
In homogeneous and isotropic turbulence
a drift from regions with increased
concentration of particles by a turbulent flow
is equiprobable in all directions, and the fluid pressure
and temperature fluctuations are not correlated with the
velocity fluctuations.

In a temperature-stratified turbulence with a non-zero
mean temperature gradient, the fluid temperature
and velocity fluctuations are correlated.
Fluctuations of temperature result in the fluid pressure fluctuations.
Increase of the fluid pressure fluctuations
is accompanied by an accumulation of particles, so that
the direction of the mean flux of particles
coincides with the turbulent heat flux, toward the
minimum of the mean temperature. This causes formation
of large-scale inhomogeneous distributions
of inertial particles.

Let us discuss the phenomenon of turbulent thermal diffusion in more detail.
Motion of inertial particles with the sizes which are much smaller
than the fluid viscous scale and their material density, $\rho_p$, is much larger
than the fluid density, $\rho$, is determined by the following equation:
\begin{eqnarray}
{d {\bm v} \over d t} =- {{\bm v}-{\bm u} \over \tau_p} + {\bm g} ,
\label{C6}
\end{eqnarray}
where ${\bm v}$ is the particle velocity, ${\bm u}$ is the fluid velocity
and  ${\bm g}$ is the acceleration of gravity.
The solution of Eq.~(\ref{C6}) for small Stokes time
is obtained by iterations \citep{M87}:
\begin{eqnarray}
{\bm v} = {\bm u} - \tau_p \, \left[{\partial {\bm u} \over \partial t}
+ ({\bm u} {\bf \cdot} \bec{\nabla}) {\bm u} \right] + \tau_p \,{\bm g} + {\rm O}({\rm St}^2),
\label{C7}
\end{eqnarray}
where ${\bm W}_g=\tau_p {\bm g}$ is the terminal fall velocity of particles
caused by the gravity field.
For large Reynolds numbers, using Eq.~(\ref{C7}) we obtain \citep{EKR96}:
\begin{eqnarray}
\bec\nabla {\bf \cdot} \, {\bm v} = \bec\nabla {\bf \cdot} \, {\bm u}
+ {\tau_p \over \rho}  \,\Delta p  + {\rm O}({\rm St}^2),
\label{C8}
\end{eqnarray}
where $p$ is the fluid pressure. This implies that the particle
velocity field is compressible even in incompressible fluid velocity
field due to the inertia effects.
The instantaneous number density, $n_p(t,{\bm r})$, of inertial particles
in a turbulent flow is determined by the following equation \citep{CH43,AP81}:
\begin{eqnarray}
{\partial n_p \over \partial t} + \bec{\nabla} {\bf \cdot}
\, (n_p {\bm v}) = D \Delta n_p ,
\label{C1}
\end{eqnarray}
where $D$ is the coefficient of Brownian
diffusion of particles. For large
P\'eclet numbers, ${\rm Pe} \equiv \ell_0\, u_0/D \gg 1$,
when molecular diffusion of particles in Eq.~(\ref{C1})
can be neglected, we get $\bec\nabla {\bf \cdot} \, {\bm v}
\propto - d\ln n_p / d t$.
Combining this equation with Eq.~(\ref{C8})
we obtain that
$d n_p / d t \propto - n_p \, (\tau_p /\rho) \,\Delta p > 0$.
This implies that in the regions with maximum fluid pressure
fluctuations, where $\Delta p < 0$, there is accumulation of
inertial particles, $d n_p / d t > 0$.
In a stratified turbulence, the fluid velocity
fluctuations are correlated with the fluid temperature and
pressure fluctuations due to a non-zero turbulent heat flux.
This causes a nondiffusive particle turbulent flux towards the regions with
the minimum of the mean fluid temperature.
This phenomenon results in the large-scale particle clustering
in a temperature stratified turbulence.

To investigate the formation of
large-scale inhomogeneous structures in particle spatial distribution,
we apply a mean-field approach and use the Reynolds averaging.
In particular, we average Eq.~(\ref{C1}) over the statistics of turbulent velocity
field to obtain an equation for the mean number density of
particles $N=\langle n_p \rangle$:
\begin{eqnarray}
{\partial N \over \partial t} + \bec\nabla {\bf
\cdot} \, \left[N {\bm W}_g + \langle n \, {\bm v} \rangle -
D \,\bec\nabla N \right] =0 \;,
\label{C2}
\end{eqnarray}
where $\langle ... \rangle$ denotes ensemble averaging.
We assumed here for simplicity vanishing mean fluid velocity.
The turbulent flux of particles, $\langle n \, {\bm v} \rangle$,
in a temperature stratified turbulence has been determined
using different analytical methods,
i.e., the dimensional analysis, the quasi-linear approach,
the path-integral approach, the spectral $\tau$ approach,
the functional multi-scale turbulence approach, etc
(see \citep{EKR96,EKR97,EKR98,EKR00,EKRS00,EKRS01,PM02,RE05}).
The turbulent flux of particles is given by the following expression:
\begin{eqnarray}
\langle n \, {\bm v} \rangle = N \, {\bm V}^{\rm eff} - {\bm D_T} \,
\bec\nabla N ,
\label{C3}
\end{eqnarray}
where ${\bm V}^{\rm eff}$ is the effective
velocity of particles and ${\bm D_T}$ is the turbulent diffusion tensor of particles.
The term $- {\bm D_T} \, \bec\nabla N$, in the flux of particles
is caused by turbulent diffusion:
\begin{eqnarray}
{\bm D_T} \, \bec\nabla N \equiv \langle \tau \, v_i \, v_j \rangle \, \nabla_j N
\approx \tau_0 \, \langle u_i \, u_j \rangle \, \nabla_j N .
\label{C5}
\end{eqnarray}
For large Peclet numbers, ${\rm Pe} \equiv \ell_0\, u_0/D \gg 1$,
the turbulent diffusion tensor is
\begin{eqnarray}
{\bm D_T} = \tau_0 \, \langle u_i \, u_j \rangle,
\label{CC5}
\end{eqnarray}
where $\tau_0=\ell_0/u_0$. To derive Eqs.~(\ref{C5}) and~(\ref{CC5})
we took into account that for small
Stokes numbers, ${\rm St} \ll 1$, the particle velocity
weakly deviates from the fluid velocity, ${\bm v} = {\bm u} + O({\rm St})$.
For an isotropic turbulence the Reynolds stress, $\langle u_i \, u_j \rangle$,
is given by $\langle u_i \, u_j \rangle=(1/3) \,
\langle {\bm u}^2 \rangle \, \delta_{ij}$. Substituting this equation in
Eq.~(\ref{CC5}), we obtain the expression for
the turbulent diffusion tensor: ${\bm D_T}=D_T \delta_{ij}$, where $D_T= \tau_0 \, \langle {\bm u}^2 \rangle/3$ is the turbulent diffusion coefficient. For small Peclet numbers and large Reynolds numbers, the turbulent diffusion coefficient is $D_T= {\rm Pe} \, \tau_0 \,\langle {\bm u}^2 \rangle/12$ (see, e.g., Appendix A in \cite{SSEKR09}).

The first term, $N \, {\bm V}^{\rm eff}$, in Eq.~(\ref{C3})
determines the contribution to the turbulent flux of particles caused by
turbulent thermal diffusion in a stratified turbulence, where the effective
velocity, ${\bm V}^{\rm eff}$, of inertial particles is \cite{EKR96,EKR97}
\begin{eqnarray}
{\bm V}^{\rm eff} = - \langle \tau \, {\bm v} \,
(\bec\nabla \cdot {\bm v}) \rangle = - \alpha D_T {{\bm \nabla}T \over T} .
\label{C4}
\end{eqnarray}
Here $T$ is the mean fluid temperature and $\alpha$ is the coefficient of the turbulent thermal diffusion. For non-inertial particles or gaseous admixtures the parameter $\alpha = 1$, while for inertial particles the parameter $\alpha$ is a function of the Reynolds and Stokes numbers,
\begin{eqnarray}
\alpha &=& 1 + {2 W_g \, L_P \, \ln({\rm Re})  \over 3 D_T} \, F({\rm Re}, d)
\nonumber\\
&=& 1 + {{\rm St} \, \ln({\rm Re}) \over {\rm Re}^{1/4}} \, \left({L_{\rm eff} \over \ell_0}\right)\, F({\rm Re}, d),
\label{B9}
\end{eqnarray}
(see \cite{EKR98,EKR00,EKR13}), where  $L_P^{-1} = |{\bm \nabla} P|/P$ is the inverse scale of the mean fluid pressure variations, $L_{\rm eff} = 2 c_s^2 \tau_\eta^{3/2}/3 \nu^{1/2}$ is the effective length scale and $c_s$ is the
sound speed. When the particle diameter $d \geq d_{\rm cr}$,  the function $F({\rm Re}, d)$  is given by $F({\rm Re}, d) = 1 - 3 \ln(d / d_{\rm cr}) / \ln({\rm Re})$,
where the critical particle diameter is $d_{\rm cr} = 2 \ell_{\eta} (\rho / \rho_{p})^{1/2}$, $\rho_p$ is the material density of a particle, and $\ell_{\eta}$ is the Kolmogorov viscous scale of turbulence.
When the particle diameter  $d < d_{\rm cr}$, the function $F({\rm Re}, d) = 1$.

The effective particle velocity of inertial particles is directed
opposite to the mean temperature gradient as well as the mean heat flux, i.e.,
towards to the mean temperature minimum. This causes accumulation
of particles in this region. This effect is called turbulent thermal diffusion
because the expression for the turbulent flux $N \, {\bm V}^{\rm eff}= - N \kappa_{_{T}} {\bm \nabla}T$ with the coefficient $\kappa_{_{T}}=\alpha D_T/T$ is similar to the formula for molecular flux caused by the molecular thermal diffusion. These two effects are of statistical nature, whereby particle turbulent flux caused by turbulent thermal diffusion arises from averaging over statistics of turbulent velocity field, while the molecular thermal diffusion flux arises from solving the Boltzmann kinetic equation.

The expression for the effective velocity of inertial particles due to turbulent thermal diffusion has been derived only for small Stokes numbers
and for a weak stratification, $\ell_0 \, |\bec{\nabla} T| / T \ll 1$
(see \citep{EKR96,EKR97,EKR98,EKR00,EKRS00,EKRS01,SSEKR09}).
In the next Section we develop the theory of this effect
for arbitrary stratifications and Stokes numbers.

\section{Theory for arbitrary stratifications and Stokes numbers}
\label{Sect.III}

\subsection{Model of a turbulent particle velocity field}

In this section we discuss a model for the second moments, $\langle v_i({\bm k}) \, v_j(-{\bm k}) \rangle$, of a particle velocity field in a low-Mach-number homogeneous stratified turbulence with  arbitrary gradients of the mean temperature and arbitrary Stokes numbers.
In anelastic approximation, div ${\bm u} = {\bm \lambda} \cdot {\bm u}$, where ${\bm \lambda}=-{\bm \nabla}\rho/\rho$ and $\rho$ is the mean fluid density. The second moments, $\langle u_i({\bm k}) \, u_j(-{\bm k}) \rangle$, of a fluid velocity field in the anelastic approximation in a Fourier space in a homogeneous turbulence with arbitrary gradients of the mean temperature have the following form:
\begin{eqnarray}
&&\langle u_i({\bm k}) \, u_j(-{\bm k}) \rangle = {\langle {\bm u}^2 \rangle \, E(k) \over 8 \pi (k^2 + \lambda^2)} \biggl[\delta_{ij} - {k_i \, k_j \over k^2} + {i \over k^2} \, \big(\lambda_i \, k_j
\nonumber\\
&& \quad - \lambda_j \, k_i\big) + {\lambda^2 \over k^2} \left(\delta_{ij} - {\lambda_i \, \lambda_j \over \lambda^2}\right)\biggr] ,
\label{B1}
\end{eqnarray}
(see derivation of this equation in Appendix~\ref{Appendix-A}),
where $\delta_{ij}$ is the Kronecker tensor,
$E(k)$ is the energy spectrum function, $E(k) = (2/3) \, k_0^{-1} \,  (k / k_{0})^{-5/3}$ with $k_0 \leq k \leq k_{\nu}$.  This energy spectrum function corresponds to the Kolmogorov turbulence, the wave number $k_{0} = 1 / \ell_0$, the wave number $k_{\eta}=\ell_{\eta}^{-1}$, and $\ell_{\eta} = \ell_0 {\rm Re}^{-3/4}$ is the Kolmogorov (viscous) scale.
Equation~(\ref{B1}) follows also from the symmetry arguments (see, e.g., Appendix~E in \cite{EKR95}). The first two terms in the quadratic brackets of Eq.~(\ref{B1}) determine the incompressible, isotropic and homogeneous part of turbulence, while the other terms depend on ${\bm \lambda}$ and correspond to the anelastic approximation for arbitrary gradients of the fluid density.

Now we generalize the model~(\ref{B1}) to the case of particle velocity field
with arbitrary Stokes numbers and for turbulence with arbitrary mean temperature gradient.
In particular, we assume that the second moments, $\langle v_i({\bm k}) \, v_j(-{\bm k}) \rangle$, of a particle velocity field in the anelastic approximation
in a Fourier space in a homogeneous stratified turbulence have the following form:
\begin{eqnarray}
&&\langle v_i({\bm k}) \, v_j(-{\bm k}) \rangle = {\langle {\bm v}^2 \rangle \, E(k) \over 8 \pi [k^2 + (B\lambda)^2]} \biggl[\delta_{ij} - {k_i \, k_j \over k^2} + {i A \over k^2} \, \big(\lambda_i \, k_j
\nonumber\\
&& \quad - \lambda_j \, k_i\big) + {(B\lambda)^2 \over k^2} \left(\delta_{ij} - {\lambda_i \, \lambda_j \over \lambda^2}\right)\biggr] ,
\label{B2}
\end{eqnarray}
where we introduced two free parameters, $A$ and $B$, that are the functions of the Reynolds and Stokes numbers to be determined in the next section.

\subsection{The effective velocity of inertial particles}

\begin{figure}
\centering
\includegraphics[width=8.5cm]{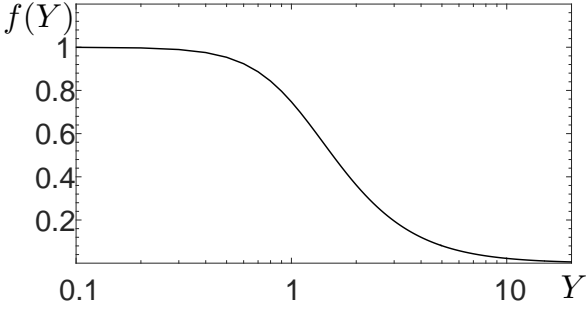}
\caption{\label{Fig1} The function $f(Y)$.}
\end{figure}

Using the model of a turbulent particle velocity field given by Eq.~(\ref{B2}), we determine the effective velocity of particles,
\begin{eqnarray}
{\bm V}^{\rm eff} = -\langle \tau {\bm v} \, {\rm div} {\bm v} \rangle = i \int \tau(k) \, k_j \langle v_i({\bm k}) \, v_j(-{\bm k}) \rangle \, d{\bm k} ,
\nonumber\\
\label{B21}
\end{eqnarray}
where $\tau(k) = 2 \tau_0 (k / k_{0})^{-2/3}$ is the scale-dependent turbulent time that corresponds to the Kolmogorov turbulence.
The details of the calculations of the particle effective velocity are given in Appendix~\ref{Appendix-B}), and the final expression for ${\bm V}^{\rm eff}$ has the following form:
\begin{eqnarray}
{\bm V}^{\rm eff} = -  A \, D_T \, f\left[\left(B \,\ell_0 \lambda \right)^{2/3}\right] \, {\bm \lambda} ,
\label{B3}
\end{eqnarray}
where $D_T=\tau_0 \, \langle {\bm v}^2 \rangle/3$,
$\lambda=|{\bm \lambda}|$, and the function $f(Y)$ is given by the following expression:
\begin{eqnarray}
f(Y) &=&  {2 \over \sqrt{3} \, Y^2} \, \biggl[{\pi \over 6} + \arctan \biggl({2 Y -1 \over \sqrt{3}}\biggr)
\nonumber\\
&& - {\sqrt{3}\over 6} \, \ln {\left(1+ Y\right)^{3} \over 1+Y^{3}} \biggr] .
\label{B4}
\end{eqnarray}
For small $Y \ll 1$, the function $f(Y)=1-Y/4 + O(Y^2)$, while for large $Y \gg 1$, this function $f(Y)=4 \pi / \left(3^{3/2} Y^2\right) + O(Y^{-3})$, and for arbitrary values of the argument $Y$ the function $f(Y)$ is shown in Fig.~\ref{Fig1}.

The equation of state for a perfect gas yields:
\begin{eqnarray}
{{\bm \nabla}\rho \over \rho} = {{\bm \nabla}P \over P} - {{\bm \nabla}T \over T} ,
\label{B8}
\end{eqnarray}
where $P$ and $T$ are the mean fluid pressure and temperature, respectively.
We assume for simplicity that the gradient of the mean fluid pressure vanishes, ${\bm \nabla}P=0$. In this case Eq.~(\ref{B3}) can be rewritten in the following form:
\begin{eqnarray}
{\bm V}^{\rm eff} = -  A \, D_T \, f\left[\big(B \,\delta_{_{T}} \big)^{2/3}\right] \, {{\bm \nabla}T \over T} ,
\label{B23}
\end{eqnarray}
where the dimensionless parameter $\delta_{_{T}}$ is defined as
\begin{eqnarray}
\delta_{_{T}} = \ell_0 {|\bec{\nabla} T| \over T} .
\label{B7}
\end{eqnarray}
It follows from Eq.~(\ref{B23}) that for a weak stratification, $B \, \delta_{_{T}} \ll 1$, the effective particle velocity is given by the following expression:
\begin{eqnarray}
{\bm V}^{\rm eff} = -  A \, D_T \, \left[1-{1\over 4} \left(B \, \delta_{_{T}} \right)^{2/3}\right] \, {{\bm \nabla}T \over T} ,
\label{B5}
\end{eqnarray}
while for a strong stratification, $B \, \delta_{_{T}} \gg 1$, the effective particle velocity is given by the following formula:
\begin{eqnarray}
{\bm V}^{\rm eff} = - {4 \pi A \, D_T \, \over 3^{3/2}} \,  \left(B \, \delta_{_{T}} \right)^{- 4/3} \, {{\bm \nabla}T \over T} .
\label{B6}
\end{eqnarray}

To determine the function $A$, we compare
Eq.~(\ref{B5}) with Eq.~(\ref{C4}) for the particle effective velocity, ${\bm V}^{\rm eff}$, derived for a small Stokes number, ${\rm St} \ll 1$, and a weak stratification, $\delta_{_{T}} \ll 1$ (see \cite{EKR96,EKR97,EKR00,EKRS00,EKRS01,EKR13}).
This comparison shows that $A = \alpha$, where the function $\alpha$ is determined by Eq.~(\ref{B9}).
Therefore, the expression for the effective particle velocity can be written as follows:
\begin{eqnarray}
{\bm V}^{\rm eff} &=& -  {2 D_T \, \alpha \over \sqrt{3} \, \big(B \, \delta_{_{T}} \big)^{4/3}}  \, \biggl[{\pi \over 6} + \arctan \biggl({2 \big(B \, \delta_{_{T}} \big)^{2/3} -1 \over \sqrt{3}}\biggr)
\nonumber\\
&& - {\sqrt{3} \over 6}\, \ln {\left[1+ \big(B \, \delta_{_{T}} \big)^{2/3}\right]^{3} \over 1+\big(B \, \delta_{_{T}} \big)^{2}} \biggr] \, {{\bm \nabla}T \over T} ,
\label{A7}
\end{eqnarray}
where we used Eq.~(\ref{B4}).
To determine the second function $B$, we assume that
\begin{eqnarray}
B=\alpha^\beta \, \varphi({\rm St},{\rm Re}) ,
\label{B20}
\end{eqnarray}
where the exponent $\beta$ should be larger than $3/4$. Indeed, when St $\to \infty$, the effective velocity should vanish, that occurs when $\beta>3/4$ [see Eq.~(\ref{B6})]. We will see in the next section that a good agreement of the results of laboratory experiments with the theoretical results is achieved when $\beta=1$ (see Sect.~\ref{Sect.III-D}).
The ratio $B/\alpha$ is a free parameter in the theory to be determined in our laboratory experiments and atmospheric observations (see Sect.~\ref{Sect.IV}).

\begin{figure}
\centering
\includegraphics[width=8.7cm]{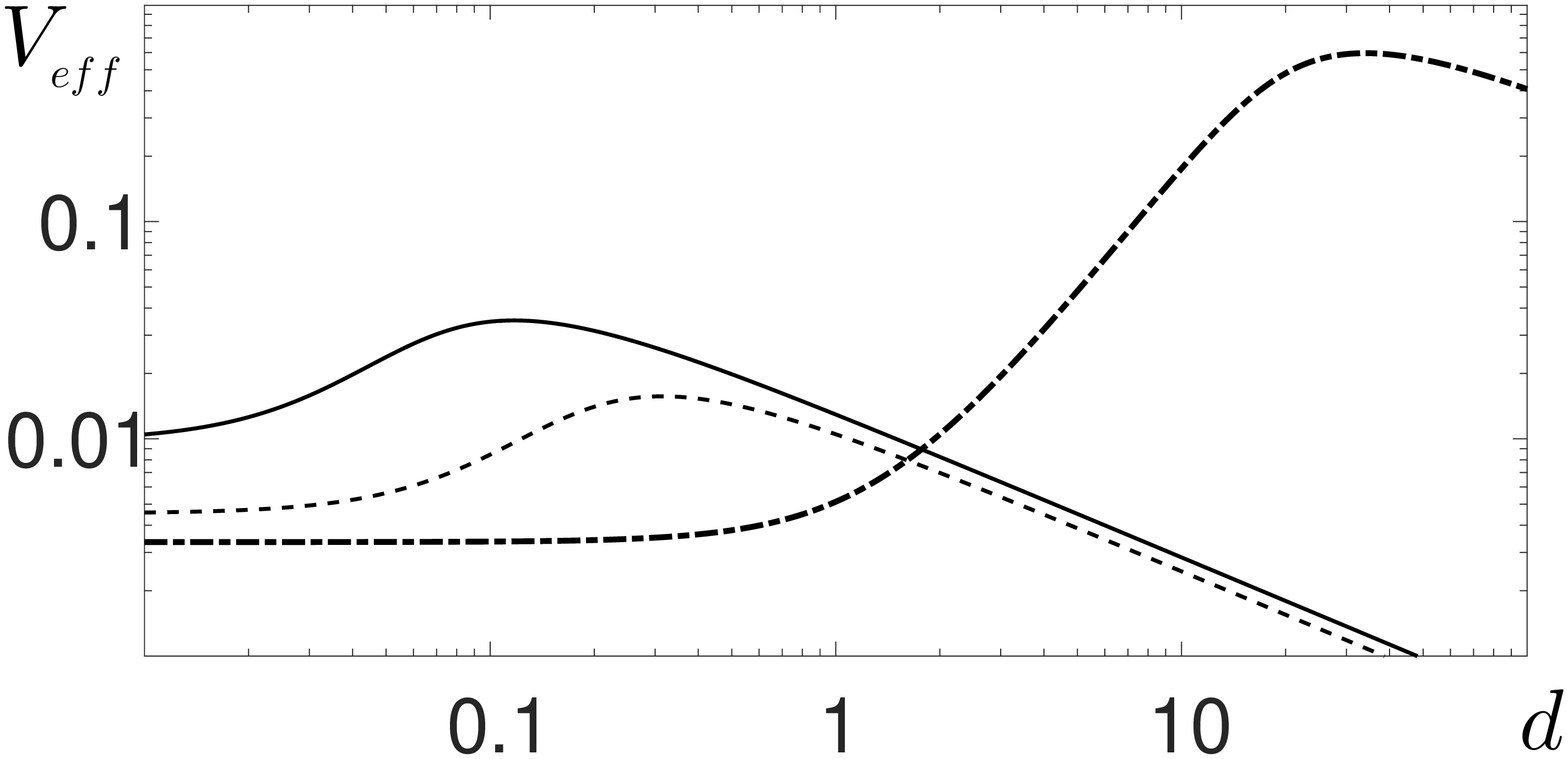}
\caption{\label{Fig2} Effective velocity $V^{\rm eff}$ measured in the units of the r.m.s turbulent vertical velocity, $u_z^{(\rm rms)}$,  versus the particle diameter $d$ ($\mu$m) for atmospheric conditions where the parameter $B/\alpha=1$ (dashed-dotted) and for laboratory experiments conditions: the oscillating grid turbulence where the parameter $B/\alpha=30$   (solid) and the multi-fan produced turbulence where the parameter $B/\alpha=18$ (dashed).}
\end{figure}

The effective velocity $V^{\rm eff}$ versus the particle diameter $d$ for conditions pertinent
to our laboratory experiments and the atmospheric turbulence is shown in Fig.~\ref{Fig2}.
Note that the characteristic Reynolds numbers
based on the turbulent integral scale in the atmospheric turbulence vary from $10^6$ to $10^7$, and in our laboratory experiments the Reynolds numbers vary from $10^2$ to $10^3$.
It follows from Fig.~\ref{Fig2} that the ratio of the effective velocity of
inertial particles to the characteristic vertical turbulent velocity for large
Reynolds numbers is less than 1.

\subsection{Turbulent diffusion of inertial particles}

Equation~(\ref{B2})  also allows us to determine the turbulent diffusion tensor for particles:
\begin{eqnarray}
D_{ij}^T = \int \tau(k) \langle v_i({\bm k}) \, v_j(-{\bm k}) \rangle \,d {\bm k} .
\label{24}
\end{eqnarray}
Using Eqs.~(\ref{B25})-(\ref{A6}) given in Appendix~\ref{Appendix-B}, we obtain:
\begin{eqnarray}
D_{ij}^T &=& {3D_T \over 2}\,\biggl\{\delta_{ij} \, \left[1 - {2 \over 3} f\left[\big(B \, \delta_{_{T}} \big)^{2/3}\right]\right]
\nonumber\\
&& - {\lambda_i \, \lambda_j \over \lambda^2} \, \biggl[1 - 2 f\left[\big(B \,\delta_{_{T}} \big)^{2/3}\right]\biggr]\biggr\} .
\label{B10}
\end{eqnarray}
It follows from Eq.~(\ref{B10}) that for a weak stratification, $B \, \delta_{_{T}} \ll 1$, the turbulent diffusion tensor for particles is given by the following expression:
\begin{eqnarray}
D_{ij}^T &=& {D_T \over 2}\,\biggl[\delta_{ij} + 3 {\lambda_i \, \lambda_j \over \lambda^2}
\nonumber\\
&& + {1 \over 2} \, \left(\delta_{ij} - 3 {\lambda_i \, \lambda_j \over \lambda^2} \right) \, \left(B \, \delta_{_{T}} \right)^{2/3}\biggr],
\label{B11}
\end{eqnarray}
while for a strong stratification, $B \, \delta_{_{T}} \gg 1$, it is given by the following formula:
\begin{eqnarray}
D_{ij}^T &=& {3 D_T \over 2}\,\biggl[\delta_{ij} - {\lambda_i \, \lambda_j \over \lambda^2}
\nonumber\\
&& + {8 \pi \over 3^{5/2}} \, \left(\delta_{ij} - 3 {\lambda_i \, \lambda_j \over \lambda^2} \right) \, \left(B \, \delta_{_{T}} \right)^{-4/3} \biggr] .
\label{B12}
\end{eqnarray}
Equations~(\ref{B10})-(\ref{B12}) show that turbulent diffusion tensor for particles is anisotropic for stratified turbulence.

\subsection{Effective coefficient of turbulent thermal diffusion}
\label{Sect.III-D}

An equation for the mean number density,
$N$, of inertial particles reads:
\begin{eqnarray}
{\partial N \over \partial t} + {\bm \nabla} {\bf
\cdot} \, \left[N \, ({\bm W}_g + {\bm V}^{\rm eff})
- (D+D_T) \, {\bm \nabla} N \right] =0 ,
\nonumber\\
\label{B13}
\end{eqnarray}
(see \citep{EKR96,EKR97,EKR98,EKR00,EKRS00,EKRS01}), where we took into
account the effect of gravity for inertial particles and molecular Brownian diffusion. However, for simplicity we neglected the effects of stratification and particle inertia on the particle turbulent diffusion coefficient.
The steady state solution of Eq.~(\ref{B13}) reads:
\begin{eqnarray}
{{\bm \nabla} N \over N} = {{\bm W}_g + {\bm V}^{\rm eff} \over D+D_T}.
\label{B14}
\end{eqnarray}
This implies that the dimensionless parameter, $\delta_{_{N}} \equiv \ell_0 |\bec{\nabla} N| / N$, characterising variations of the mean particle number density, is given by the following formula:
\begin{eqnarray}
\delta_{_{N}} = {\ell_0 \over 1 + D/D_T}  \left|\alpha \,  f\left[\big(B \,\delta_{_{T}} \big)^{2/3}\right] \, {{\bm \nabla}T \over T} - {\tau_p {\bm g} \over D_T} \right| .
\label{B15}
\end{eqnarray}
When the gradient of the mean temperature, $\bec{\nabla} T$, is directed along (or opposite to) the vertical direction, Eq.~(\ref{B15}) yields the following ratio:
\begin{eqnarray}
{\alpha^{\rm eff} \over \alpha} &=& {1 \over 1 + D/D_T}  \biggl|f\left[\big(B \,\delta_{_{T}} \big)^{2/3}\right]
\nonumber\\
&& +  {\tau_p \, g \, \ell_0 \over \alpha \,\delta_{_{T}} \, D_T} \, {\rm sgn}(\nabla_z T) \biggr| ,
\label{B16}
\end{eqnarray}
where we introduced a new parameter $\alpha^{\rm eff}\equiv\delta_{_{N}}/\delta_{_{T}}$.
It follows from Eq.~(\ref{B16}) that for a weak stratification,
$B \, \delta_{_{T}} \ll 1$, the ratio
$\alpha^{\rm eff} / \alpha$ is given by the following formula:
\begin{eqnarray}
{\alpha^{\rm eff} \over \alpha} &=& {1 \over 1 + D/D_T}  \biggl|1-{1\over 4} \left(B \, \delta_{_{T}} \right)^{2/3}
\nonumber\\
&& +  {\tau_p \, g \, \ell_0 \over \alpha \,\delta_{_{T}} \, D_T} \, {\rm sgn}(\nabla_z T) \biggr| ,
\label{B17}
\end{eqnarray}
while for a strong stratification, $B \, \delta_{_{T}} \gg 1$, it is given by the following expression:
\begin{eqnarray}
{\alpha^{\rm eff} \over \alpha} &=& {1 \over 1 + D/D_T}  \biggl|{4 \pi \over 3^{3/2}} \,  \left(B \, \delta_{_{T}} \right)^{- 4/3}
\nonumber\\
&& +  {\tau_p \, g \, \ell_0 \over \alpha \,\delta_{_{T}} \, D_T} \, {\rm sgn}(\nabla_z T) \biggr| .
\label{B18}
\end{eqnarray}

\begin{figure}
\centering
\includegraphics[width=8.5cm]{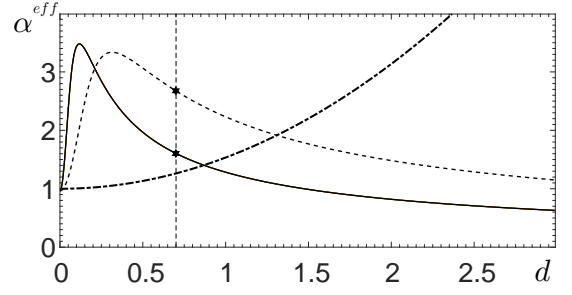}
\caption{\label{Fig3} Dependencies of the effective turbulent thermal diffusion coefficient $\alpha^{\rm eff}$ versus the particle diameter $d$ ($\mu$m) for atmospheric conditions where the parameter $B/\alpha=1$ (dashed-dotted), and for laboratory experiments with oscillating grid turbulence where the parameter $B/\alpha=30$ (solid), and multi-fan produced turbulence where the parameter $B/\alpha=18$ (dashed). Vertical line corresponds to the particle diameter $d=0.7$ $\mu$m used in the experiments (stars are values obtained from measurements).}
\end{figure}

The dependencies of the effective turbulent thermal diffusion coefficient $\alpha^{\rm eff}$ versus the particle diameter $d$ for conditions pertinent for atmospheric turbulence and laboratory experiments are shown in Fig.~\ref{Fig3}. To isolate turbulent thermal diffusion from other effects hereafter we do not take into account the gravity effect.
Inspection of Fig.~\ref{Fig3} shows that for the conditions pertinent for our laboratory experiments the effective turbulent thermal diffusion coefficient $\alpha^{\rm eff}$ increases for very small particle size $d$, reaches the maximum at small $d$ and slowly decreases for larger particle size. In the atmospheric turbulence the effective turbulent thermal diffusion coefficient $\alpha^{\rm eff}$ behaves in a similar way, except for $\alpha^{\rm eff}$ reaches the maximum at much larger particle size $d$.

\section{Validation of theory in laboratory experiments and atmospheric turbulence}
\label{Sect.IV}

To validate the theory of turbulent thermal
diffusion in strongly stratified
turbulent flows we perform laboratory experiments
in different set-ups: in the oscillating grid turbulence
(see, e.g., \cite{BEE04,EEKR04,EEKMR06,EKR10,EEKR11,EEKR13})
and in the multi-fan produced turbulence (see, e.g., \cite{EEKR06}).
We also validate the theory of turbulent thermal diffusion against
data of meteorological observations (see, e.g., \cite{SSEKR09}).

\subsection{Particles in the oscillating grid turbulence}

In this section we describe very briefly the experimental set-up
and measurement facilities in the oscillating grid turbulence.
The details of the experimental set-up and measurements
in the oscillating grid turbulence
can be found in \cite{EEKMR06,EKR10,EEKR11,EEKR13}.
The experiments in stratified
turbulence have been conducted in rectangular
chamber with dimensions $26 \times 58 \times 26$
cm$^3$ in air flow.
In the experiments turbulence is produced by
two oscillating vertically oriented grids
with bars arranged in a square
array. The grids are parallel to the side
walls and positioned at a distance of two grid meshes from
the chamber walls. They
are operated at the same
amplitude, at a random phase and at
the same frequency up to $10.5$ Hz.

A vertical mean temperature gradient in the
turbulent air flow was formed by attaching two
aluminium heat exchangers to the bottom and top
walls of the test section which allowed us to
form a mean temperature gradient up to 1.15 K/cm
at a mean temperature of about 308 K when the
frequency of the grid oscillations $f = 10.5$ Hz.
To improve heat transfer in the boundary layers
at the bottom and top walls we used heat exchangers with
rectangular fins $0.3 \times 0.3 \times 1.5$ cm$^3$.
The temperature field was measured with a temperature
probe equipped with a vertical array of 12 E-thermocouples in the
central part of the chamber in many locations.

The velocity fields were measured using a
Stereoscopic Particle Image  Velocimetry (PIV)
with LaVision Flow Master III system.
We obtain velocity maps in the central region of
the flow in the cross-section perpendicular to the
grids and parallel to a front view plane.
An incense smoke with sub-micron particles
($\rho_p / \rho \sim 10^3)$,  was used as a
tracer for the PIV measurements. Smoke was
produced by high temperature sublimation of solid
incense grains. These particles
have an approximately spherical shape and the mean
diameter of 0.7 $\mu$m.

We determined the mean  and the
r.m.s.~velocities, two-point correlation
functions and an integral scale of turbulence
from the measured velocity fields. Series of 520
pairs of images acquired with a frequency of 2 Hz,
were stored for calculating velocity maps and
for ensemble and spatial averaging of turbulence
characteristics.
We measured velocity in a flow domain $32.8 \times 24.8$ cm$^2$ with a
spatial resolution of $0.24$ mm/pixel.
The mean and r.m.s. velocities for
every point of a velocity map were calculated by
averaging over 520 independent velocity maps, and then
they were spatially averaged over the central flow region.
An integral scale of turbulence, $\ell$, was
determined from the two-point correlation
functions of the velocity field.
The characteristic turbulence time in the
experiments is much smaller than the time during which the
velocity fields are measured $(260$ s).
We performed experiments for different temperature difference,
$\Delta T$, between the top and bottom plates.

Spatial distributions for 0.7 $\mu$m and 10 $\mu$m particles
were determined by Particle
Image Velocimetry (PIV) system using the effect of Mie
light scattering by particles in the flow \cite{BEE04,EEKR04,EEKMR06}.
In order to characterize the spatial distribution of particle number
density in the non-isothermal flow, the
distribution of the scattered light intensity measured in the
isothermal case was used for the normalization of the scattered
light intensity obtained in a non-isothermal flow under the
same conditions. The scattered light intensities in
each experiment were also normalized by corresponding scattered light
intensities averaged over the vertical coordinate.

For experimental study of turbulent thermal diffusion of inertial particles we
used  borosilicate hollow glass particles having an approximately
spherical shape, a mean diameter of $10 \, \mu$m
and the material density $ \rho_p \approx 1.4 $ g/cm$^3$.
These particles have been injected in the chamber using an air jet
in order to improve particle mixing and prevent from particle agglomeration.
We used a custom-made acoustic feeding device for injecting particles into the flow comprising an acrylic glass chamber with a size of 9 $\times$ 9 $\times$ 4 cm$^3$. Two plastic slabs inside the chamber are used as air guides to achieve optimal flow with entrained particles. Particle dispensation zone (a disk of 25 mm diameter and 5 mm thickness) is located at the bottom of the chamber. A standard woofer (oval 2 $\times$ 3.5") at a frequency of 220 Hz sways a latex membrane on which particles are loaded. A cylindrical cavity is used to contain the particles on the latex membrane. The batch of particles on the membrane should roughly fill the cavity.
When the membrane vibrates, particles are entrained into air.  Particle feeding device has a pressurized air inlet with a bellow having a 8 mm diameter tube with standard quick release connector. The entrained particles leave the chamber with a stream of air through the outlet.

The parameters in the oscillating grid turbulence are as follows: the integral length scales in the vertical and horizontal directions are $\ell_z=1.4$ cm and $\ell_y=2.2$ cm;
the characteristic turbulent vertical and horizontal velocities (the root mean square velocity) are $u_z=7.8$ cm s$^{-1}$ and $u_y=17.3$ cm s$^{-1}$; the characteristic turbulent times in these directions are $\tau_z=0.18$ s and $\tau_y=0.13$ s;
the vertical and horizontal Reynolds numbers are Re$_z\equiv \ell_z u_z/\nu=73$ and Re$_y\equiv \ell_y u_y/\nu=254$.

\begin{figure}
\centering
\includegraphics[width=9.0cm]{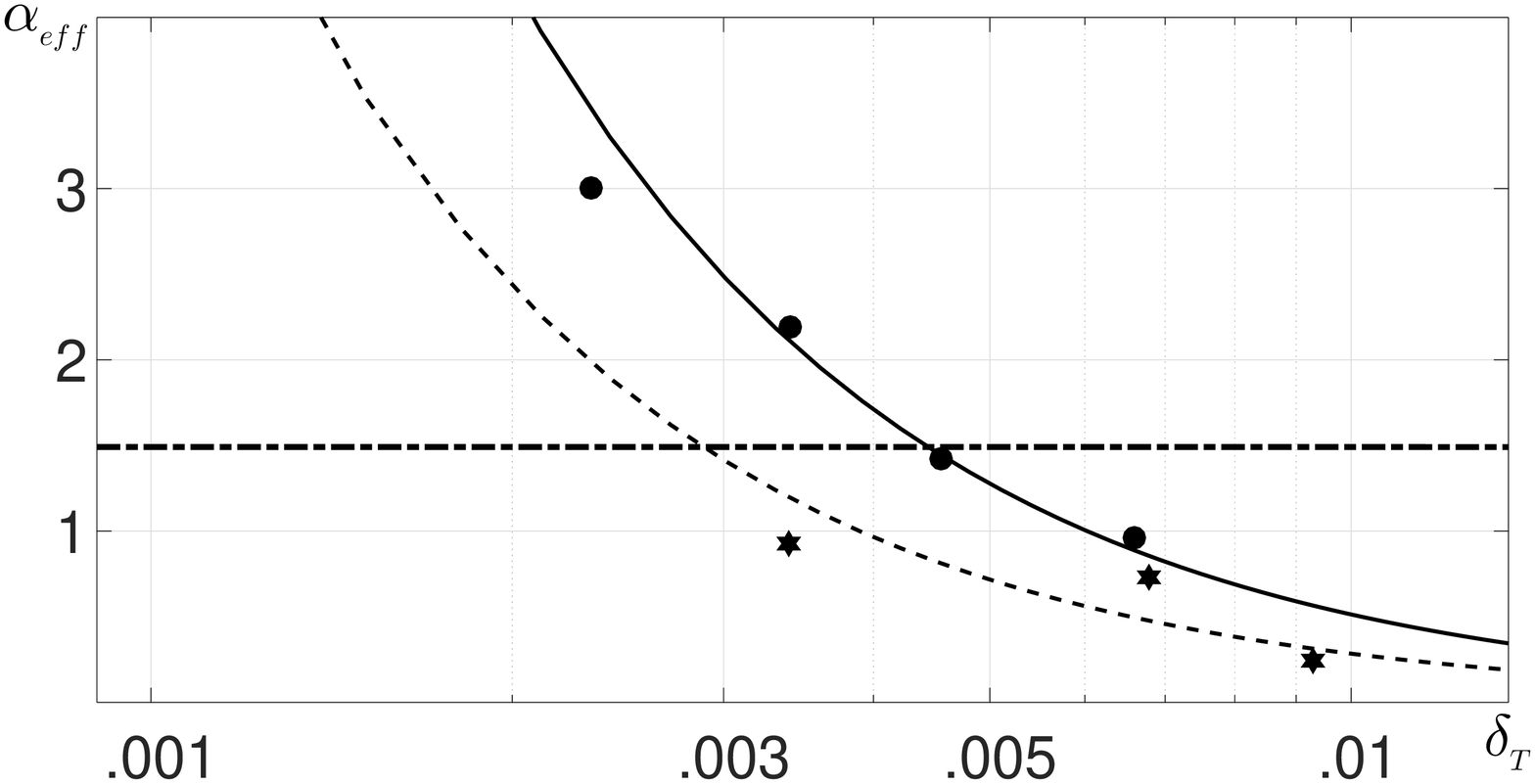}
\caption{\label{Fig4} Effective turbulent thermal diffusion coefficient $\alpha^{\rm eff}$ versus the parameter $\delta_{_{T}}$ for 0.7 $\mu$m particles with the parameter
$B/\alpha=21$ (solid) and for 10 $\mu$m particles with the parameter $B/\alpha=18$ (dashed).
The laboratory experiments with oscillating grid turbulence are performed with
0.7 $\mu$m particles (circle) and 10 $\mu$m particles (star) for different temperature difference $\Delta T$ between the top and bottom heat exchangers.
For comparison the function $\alpha^{\rm eff}(\delta_{_{T}})$ is  also shown
for atmospheric turbulence (dashed-dotted) for 1 $\mu$m particles
and the parameter $B/\alpha=1$.}
\end{figure}

In Fig.~\ref{Fig4} we show the effective turbulent thermal diffusion coefficient $\alpha^{\rm eff}$ versus the parameter $\delta_{_{T}}$ for 0.7 $\mu$m particles with the parameter $B/\alpha=21$ and for 10 $\mu$m particles with the parameter $B/\alpha=18$.
The laboratory experiments with oscillating grid turbulence are performed with
0.7 $\mu$m particles and 10 $\mu$m particles for different temperature difference $\Delta T$ between the top and bottom heat exchangers.
For comparison we also show in Fig.~\ref{Fig4} the function $\alpha^{\rm eff}(\delta_{_{T}})$
for atmospheric turbulence for 1 $\mu$m particles, where the parameter $B/\alpha=1$. The measured values of $\alpha^{\rm eff}$ are in a good agreement with theoretical predictions.

\subsection{Particles in the multi-fan produced turbulence}

In this section we describe very briefly  the experimental set-up
and measurement facilities in the multi-fan produced turbulence.
The details of the experimental set-up and measurements
in the multi-fan produced turbulence
can be found in \cite{EEKR06}.
Experiments were conducted in
a multi-fan turbulence generator that
is the perspex cube box with dimensions $40 \times 40 \times
40$ cm$^3$. It includes eight fans
with rotation frequency of up to 2800
rpm mounted in the corners of the box and facing the
center of the box.

At the top and bottom  walls of the Perspex box two
heat exchangers with rectangular $0.3 \times 0.3 \times 1.5$ cm$^3$ fins
were installed to improve heat transfer in the boundary layers
at the bottom and top walls.
The upper wall was heated up to 343 K, the bottom wall was cooled to
283 K. Two additional fans were installed
at the bottom and top walls of the chamber in order to produce a
large mean temperature gradient $(\sim 0.92$ K/cm) in the core of the flow.
The temperature was measured with a high-frequency response thermocouple
which was glued externally to a wire.
Velocity fields and particle spatial distribution were determined
using digital Particle Image Velocimetry (PIV) system (see previous
subsection). The laboratory experiments with the multi-fan produced turbulence
are performed with 0.7 $\mu$m particles for $\Delta T=50$ K.

\begin{figure}
\centering
\includegraphics[width=9.0cm]{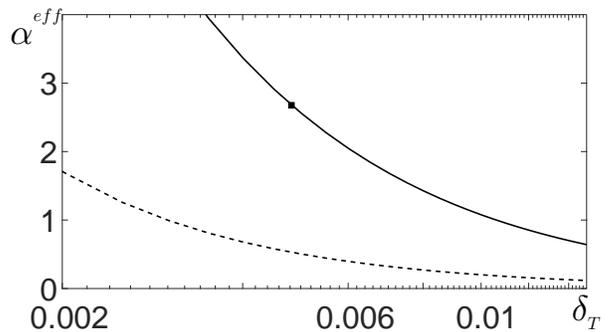}
\caption{\label{Fig5} Effective turbulent thermal diffusion coefficient $\alpha^{\rm eff}$ versus the parameter $\delta_{_{T}}$ for 0.7 $\mu$m particles (solid) and
for 10 $\mu$m particles (dashed) for the parameter $B/\alpha=18$.
The laboratory experiments with the multi-fan produced turbulence are performed with 0.7 $\mu$m particles for the temperature difference $\Delta T=50$ K between the top and bottom heat exchangers (square).}
\end{figure}

The parameters of the multi-fan produced turbulence are as follows: the integral length scales in the vertical and horizontal directions are $\ell_z=1.64$ cm and $\ell_y=1.49$ cm;
the characteristic turbulent vertical and horizontal velocities (the root mean square velocity) are $u_z=80$ cm s$^{-1}$ and $u_y=71$ cm s$^{-1}$; the characteristic turbulent times in these directions are $\tau_z=2.05 \times 10^{-2}$ s and $\tau_y=2.1\times 10^{-2}$ s;
the vertical and horizontal Reynolds numbers are Re$_z\equiv \ell_z u_z/\nu=875$ and Re$_y\equiv \ell_y u_y/\nu=705$.

Unfortunately, the region of isotropic and
homogeneous turbulence in the multi-fan produced turbulence
is not large.
The presence of 10 fans in this set-up does not allow us to perform
velocity and temperature measurements, and to obtain spatial profiles of particle
number density in many locations. This is a reason why we
performed only several experiments in the multi-fan produced turbulence.

In Fig.~\ref{Fig5} we show the effective turbulent thermal diffusion coefficient $\alpha^{\rm eff}$ versus the parameter $\delta_{_{T}}$ for 0.7 $\mu$m particles (solid line) and
for 10 $\mu$m particles (dashed line) for the parameter $B/\alpha=18$.
The measured value of $\alpha^{\rm eff}$ is in an agreement with theoretical predictions.

\begin{figure}
\centering
\includegraphics[width=9.0cm]{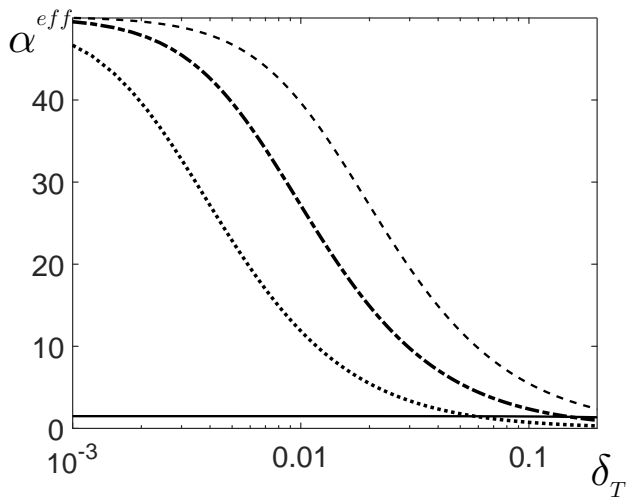}
\caption{\label{Fig6} Effective turbulent thermal diffusion coefficient $\alpha^{\rm eff}$ versus the parameter $\delta_{_{T}}$  for 1 $\mu$m particles and $B/\alpha=1$ (solid) and
for 10 $\mu$m particles and different values of the parameter $B/\alpha$:  $\; B/\alpha=1$ (dashed), $B/\alpha=2$ (dashed-dotted), $B/\alpha=5$ (dotted), and other parameters corresponds to the atmospheric turbulence conditions.}
\end{figure}

\subsection{Particles in the atmospheric turbulence}

Tropopause in the atmosphere is a well-known region with strong gradients of temperature and also with substantial amount of aerosol particles, which remain there over long time (see, e.g., \cite{BS05}).
The theory of turbulent thermal diffusion for small temperature stratifications
and small Stokes numbers has been previously applied in \cite{SSEKR09}
to the GOMOS (Global Ozone Monitoring by Occultation of Stars) aerosol observations near the tropopause in order to explain the shape of aerosol vertical profiles with elevated concentrations located almost symmetrically with respect to temperature profile.
The altitude of the GOMOS measurements is in the range from 5 km to 20 km.
Analysis of data of simultaneous observations of the vertically-resolved aerosol concentrations and mean temperature in the vicinity of the tropopause shows that the aerosol concentration and temperature profiles are often anti-correlated \cite{SSEKR09}. These observations are explained using the effect of turbulent thermal diffusion, where the turbulent flux of particles is directed towards to the mean temperature minimum.

In Fig.~\ref{Fig6} we show the effective turbulent thermal diffusion coefficient $\alpha^{\rm eff}$ obtained from the generalized theory versus the parameter $\delta_{_{T}}$  for 1 $\mu$m particles and $B/\alpha=1$, and for 10 $\mu$m particles and different values of the parameter $B/\alpha$.
It is seen in Fig.~\ref{Fig6} that $\alpha^{\rm eff}$ depends strongly on the particle size.
Note that in the atmospheric turbulent flows the parameter $\delta_{_{T}}$ varies in the range from $10^{-3}$ to $10^{-2}$ (see, e.g., \cite{SSEKR09}).
This implies that $\alpha^{\rm eff}/\alpha$ is of the order of 1 in this range of the mean temperature variations.
For example, for aerosols having the diameter of 1-3 $\mu$m,  the coefficient $\alpha^{\rm eff} \approx \alpha$  exceeds 1 when the turbulent diffusion coefficient $D_T \sim 10^4$ cm$^2$ s$^{-1}$ or less. This is in agreement with the data obtained from the GOMOS aerosol observations near the tropopause \cite{SSEKR09}.

\section{Discussion and conclusions}
\label{Sect.V}

In the present study we have investigated turbulent thermal diffusion of small inertial particles in the temperature stratified turbulence.
This effect results in the appearance of a non-diffusive turbulent flux of particles
directed towards the turbulent heat flux.

The theory of turbulent thermal diffusion has been previously
developed only for small temperature gradients
and small Stokes numbers \citep{EKR96,EKR97,EKR00,EKRS00,EKRS01,PM02,RE05}.
In the present study we have generalized the theory of turbulent thermal diffusion for arbitrary temperature gradients and Stokes numbers.
We have also performed laboratory experiments
in the oscillating grid turbulence and in the multi-fan produced turbulence
to validate the theory of turbulent thermal diffusion in strongly stratified
turbulent flows.

Turbulent flux of inertial particles caused by turbulent thermal diffusion is proportional to the product of the effective velocity of inertial particles and the mean particle number density. We have shown that the ratio of the effective velocity of
inertial particles to the characteristic vertical turbulent velocity for large
Reynolds numbers is less than 1.
We demonstrated that the effective velocity of inertial particles
increases with the increase of the Stokes numbers, reaches the maximum at small Stokes numbers
and decreases for larger Stokes numbers. In the laboratory experiments
the effective velocity of inertial particles reaches the maximum at St $=10^{-4}$,
while for the atmospheric turbulence it reaches the maximum at St $=0.05$.
The effective coefficient of turbulent thermal diffusion decreases with the mean temperature gradient. For very large Reynolds numbers this dependence on the mean temperature gradient
is very weak.
The obtained results of the laboratory experiments are in a good agreement
with the theoretical predictions.
However, the results obtained in our laboratory experiments
clearly indicate the difficulty associated
with the experimental observation of the phenomenon of turbulent thermal diffusion,
i.e., this effect is quite strong only in a certain range of values of Stokes number,
flow Reynolds number and temperature gradient.

\begin{acknowledgements}
This work has been supported by the Israel
Science Foundation governed by the Israeli
Academy of Sciences (grant No. 1210/15).
\end{acknowledgements}

\appendix
\section{\bf Model of turbulent velocity field}
\label{Appendix-A}

The anelastic condition div  ${\bm u} = {\bm u} \cdot {\bm \lambda}$
in the ${\bf k}$-space implies that $k_1^{(i)} \langle u_i({\bm k}_1) \, u_j({\bm k}_2) \rangle = 0$ and $k_2^{(j)} \langle u_i({\bm k}_1) \, u_j({\bm k}_2) \rangle = 0$, where $k_1^{(i)} = k_i + i \lambda_i$ and $k_2^{(i)} = -k_i + i \lambda_i$.
We consider the model of the turbulent velocity field in the following form:
\begin{eqnarray}
&&\langle u_i({\bm k}_1) \, u_j({\bm k}_2) \rangle =  - \langle {\bm u}^2 \rangle\, \Phi(k) \, \biggl[\delta_{ij} \left({\bm k}_1 \cdot {\bm k}_2\right)- k_1^{(i)} \, k_2^{(j)} \biggr] ,
\nonumber\\
\label{AB1}
\end{eqnarray}
where ${\bm k}_1 \cdot {\bm k}_2= -(k^2 + \lambda^2)$
and $k_1^{(i)} \, k_2^{(j)}=  - (k_i \, k_j - i \lambda_i \, k_j + i \lambda_j \, k_i + \lambda_i \, \lambda_j)$, so that
\begin{eqnarray}
&&\langle u_i({\bm k}_1) \, u_j({\bm k}_2) \rangle = \langle {\bm u}^2 \rangle \, \Phi(k) \, k^2 \biggl[\delta_{ij} - {k_i \, k_j \over k^2} + {i \over k^2} \, \big(\lambda_i \, k_j
\nonumber\\
&& \quad - \lambda_j \, k_i\big) + {\lambda^2 \over k^2} \left(\delta_{ij} - {\lambda_i \, \lambda_j \over \lambda^2}\right)\biggr] .
\label{BB1}
\end{eqnarray}
Here $\Phi(k)$ is unknown function to be determined below.
Integrating the correlation function $\langle u_i({\bm k}_1) \, u_i({\bm k}_2) \rangle$ over ${\bm k}$ we obtain
\begin{eqnarray}
&& \int \langle u_i({\bm k}_1) \, u_i({\bm k}_2) \rangle \,d {\bm k} = 2 \langle {\bm u}^2 \rangle \, \int \Phi(k)\, (k^2 + \lambda^2) \,d {\bm k}
\nonumber\\
&& = 8 \pi  \langle {\bm u}^2 \rangle \, \int \Phi(k)\, (k^2 + \lambda^2) \, k^2 \,dk .
\label{DB1}
\end{eqnarray}
On the other hand,
\begin{eqnarray}
&& \int \langle u_i({\bm k}_1) \, u_i({\bm k}_2) \rangle \,d {\bm k}
\equiv \langle {\bm u}^2 \rangle \int E(k) \,k^2 \,dk ,
\label{DB2}
\end{eqnarray}
where $E(k)$ is the spectrum function of the turbulent velocity field.
Comparing Eqs.~(\ref{DB1}) and~(\ref{DB2}), we obtain
the function $\Phi(k)$:
\begin{eqnarray}
\Phi(k)= {E(k)  \over 8 \pi (k^2 + \lambda^2)}.
\label{DB3}
\end{eqnarray}
Substituting this function into Eq.~(\ref{AB1}), we obtain Eq.~(\ref{B1}).

\section{Derivation of equation for the effective velocity}
\label{Appendix-B}

In this Appendix we derive equation for the effective velocity of particles.
Using Eqs.~(\ref{B2}) and~(\ref{B21}), we obtain:
\begin{eqnarray}
{\bm V}^{\rm eff} &=& -\langle \tau {\bm v} \, {\rm div} {\bm v} \rangle = i \int \tau(k) \, k_j \langle v_i({\bm k}) \, v_j(-{\bm k}) \rangle \, d{\bm k}
\nonumber\\
&=& - 2 D_T A \bec{\lambda} \int_0^1 {\bar\tau \,d \bar\tau \over 1 + a \bar\tau^3},
\label{A3}
\end{eqnarray}
where $a = (B \,\ell_0 \lambda)^{2}$, $\ell_0=k_0^{-1}$ and $\tau(k) = 2 \tau_0 \bar\tau(k)$.
For the integration over angles in ${\bm k}$-space in Eq.~(\ref{A3}) we used the following integrals:
\begin{eqnarray}
&&\int_{0}^{2\pi} \, d\varphi \int_{0}^{\pi} \sin \vartheta \,d\vartheta = 4 \pi .
\label{D19}\\
&&\int_{0}^{2\pi} \, d\varphi \int_{0}^{\pi} \sin \vartheta \,d\vartheta \,
{k_i \, k_j \over k^2} = {4 \pi \over 3} \, \delta_{ij} .
\label{D20}
\end{eqnarray}
Now let us calculate the integral over $\bar\tau$ in Eq.~(\ref{A3}):
\begin{eqnarray}
f(a)&=& 2 \int_0^1 {\bar\tau \,d \bar\tau \over 1 + a \bar\tau^3} = {2 \over a^{2/3}} \,  \int_0^{a^{1/3}} {X \,d X \over 1 + X^3}
\nonumber\\
&=& {2 \over a^{2/3}} \, \left[\int_0^{a^{1/3}} {\,d X \over 1 - X + X^2} -\int_0^{a^{1/3}} {\,d X \over 1 + X^3} \right],
\label{B25}
\nonumber\\
\end{eqnarray}
where $X=a^{1/3}\bar\tau$, and
\begin{eqnarray}
&& \int{ \,d X \over 1 + X^3} = {1 \over 6} \ln {(1+X)^3 \over 1 + X^3} + {1 \over \sqrt{3}}
\arctan {2X-1 \over \sqrt{3}} ,
\nonumber\\
\label{A4}\\
&& \int {\,d X \over 1 - X + X^2} = {2 \over \sqrt{3}}
\arctan {2X-1 \over \sqrt{3}} .
\label{A5}
\end{eqnarray}
Substituting Eqs.~(\ref{A4}) and~(\ref{A5}) into Eq.~(\ref{B25}) we obtain the expression for the function $f(a)$:
\begin{eqnarray}
f(a)&=& {2 \over \sqrt{3} \,  a^{2/3}} \, \biggl[{\pi \over 6} +
\arctan \left({2 a^{1/3} -1 \over \sqrt{3}} \right)
\nonumber\\
&& - {\sqrt{3}\over 6} \,  \ln {\left(1+a^{1/3}\right)^{3} \over 1+a}\biggr] .
\label{A6}
\end{eqnarray}
Therefore, the effective velocity of particles caused by turbulent thermal diffusion
is ${\bm V}^{\rm eff} = - A \, D_T \, f\left[(B \,\ell_0 \lambda)^{2}\right] \, \bec{\lambda}$.

\end{document}